\documentclass[prl,aps,twocolumn,showpacs]{revtex4}

\usepackage{graphicx}

\begin{document}

  \title{Cooperative Effect of Electron Correlation and Spin-Orbit
  Coupling on the Electronic and Magnetic Properties of Ba$_2$NaOsO$_6$} 

 \author{H. J. Xiang}
 \affiliation{Department of Chemistry, North Carolina State University, Raleigh,
 North Carolina 27695-8204}

 \author{M.-H. Whangbo}
 \thanks{Corresponding author. E-mail: mike\_whangbo@ncsu.edu}

 \affiliation{Department of Chemistry, North Carolina State University, Raleigh,
 North Carolina 27695-8204}

\date{\today}

\begin{abstract}
The electronic and magnetic properties of the cubic double perovskite
Ba$_2$NaOsO$_6$ were examined by performing first-principles density
functional theory calculations and analyzing spin-orbit coupled states
of an Os$^{7+}$ (d$^1$) ion at an octahedral crystal field. The insulating
behavior of Ba$_2$NaOsO$_6$ was shown to originate from a cooperative effect
of electron correlation and spin-orbit coupling. This cooperative
effect is responsible not only for the absence of orbital ordering in
Ba$_2$NaOsO$_6$ but also for a small magnetic moment and a weak magnetic
anisotropy in Ba$_2$NaOsO$_6$. 
\end{abstract}


\maketitle
Oxides of orbitally degenerate 3d transition metal ions at octahedral
sites exhibit rich electronic and magnetic properties arising from the
interplay between spin, orbital and charge degrees of freedom. In an
octahedral crystal field, the d-orbitals of a transition metal ion are
split into the $t_{2g}$ and $e_g$ levels. A 3d perovskite with unevenly filled
$e_g$ levels, e.g., LaMnO$_3$ with high-spin Mn$^{3+}$ (3d$^4$) ions, has a strong
tendency for orbital ordering \cite{Yin2006}. In contrast, a 3d perovskite with
unevenly filled $t_{2g}$ levels, e.g., YTiO$_3$ with Ti$^{3+}$ (3d$^1$) ions, has a
reduced tendency for orbital ordering and hence provides opportunities
to observe an intricate interplay between the spin and orbital
dynamics \cite{Ulrich2002,Schmitz2005}. It is an important issue to understand the mechanisms
that select the ground state out of numerous possible states arising
from this competition
\cite{Ulrich2002,Schmitz2005,Khaliullin2000,Pavarini2004}. 

\begin{figure}[!hbp]
  \includegraphics[width=6.0cm]{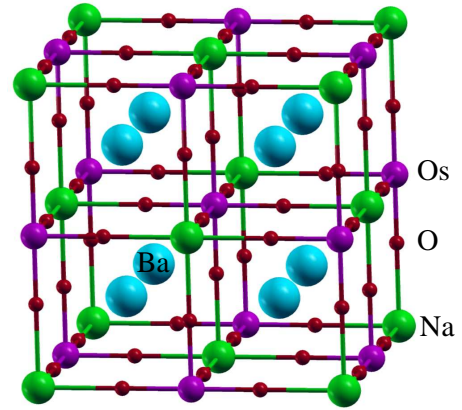}
  \caption{(Color online) Structure of cubic double perovskite Ba$_2$NaOsO$_6$.}
  \label{fig1}
\end{figure}

Most studies probing this question have
focused on 3d oxides, and much less is known about whether related 4d
and 5d oxides can exhibit similar behavior. Due to a large spatial
extension of 5d orbitals, effects of electron-correlation are weaker
in 5d oxides than in 3d oxides. However, effects of spin-orbit
coupling (SOC) are stronger in 5d oxides than in 3d oxides. Thus, 5d
oxides should exhibit a different balance between spin, orbital and
charge degrees of freedom. In this context, it is of interest to
examine the electrical and magnetic properties of the cubic double
perovskite Ba$_2$NaOsO$_6$ \cite{Sleight1962,Stitzer2002,Erickson2006}. In this 5d oxide the NaO$_6$ octahedra share
corners with the OsO$_6$ octahedra of orbitally degenerate Os$^{7+}$ (5d$^1$)
ions, and the nearest-neighbor OsO$_6$ octahedra run along the [110]
direction while the next-nearest-neighbor OsO$_6$ octahedra run along the
[100] direction (Figure~\ref{fig1}). Ba$_2$NaOsO$_6$ presents several puzzling
properties. It is an insulator despite the fact that the structure
remains cubic down to 5 K without any distortion of the OsO$_6$ octahedra
from their regular octahedral shape [8]. It is unclear what mechanism
lifts the orbital degeneracy of the Os$^{7+}$ (d$^1$) ions to make Ba$_2$NaOsO$_6$
insulating. The magnetic susceptibility of Ba$_2$NaOsO$_6$ between 75 and
200 K follows a Curie-Weiss law with a negative Weiss temperature
(i.e. $\theta \approx  -10$ K), which shows that the dominant spin exchange
interaction between Os$^{7+}$ ions is antiferromagnetic (AFM)
\cite{Stitzer2002,Erickson2006}. However, Ba$_2$NaOsO$_6$ undergoes a
ferromagnetic (FM) ordering 
below $T_C = 6.8$ K with a very low magnetic moment, i.e., $\sim$ 0.2 $\mu_B$ per
formula unit (FU) \cite{Stitzer2002,Erickson2006}.

\begin{table}
  \label{table1}
  \caption{Comparison of the zero-field magnetic moments $\mu_{exp}$  (per FU)
  with the calculated moments $\mu_{calc}$  (per FU) for the ferromagnetic state
  of Ba$_2$NaOsO$_6$ using the GGA+SOC+U method with $U_{eff} = 0.2$
  Ryd $^{a,b}$. }
  \begin{tabular}{cccccc}
    \hline
    \hline
    & $\mu_S$(Os)  & $\mu_L$(Os) & $\mu_S$ & $\mu_{calc}$ &
    $\mu_{exp}$ $^c$ \\
    \hline
	\ [111] & 0.51 & -0.35 &0.98 & 0.63 &0.19 \\
	\ [110] & 0.51 & -0.34 &0.98 & 0.64 &0.22 \\
	\ [100] & 0.52 & -0.30 &0.97 & 0.67 &0.18 \\
    \hline
    \hline
  \end{tabular}
  \\
  \begin{tabular}{l}
   $^a$ $\mu_S$(Os) and $\mu_L$(Os) are the spin and orbital magnetic 
    \\ moments calculated for the Os atom,
   respectively, and \\
   $\mu_S$   is   
   the calculated total spin magnetic moment per FU. \\
   $^b$ The moments are in units of $\mu_B$. \\ 
   $^c$ Ref.~\cite{Erickson2006}
   \end{tabular}
\end{table}

%\clearpage

\begin{figure}[!hbp]
  \includegraphics[width=6.0cm]{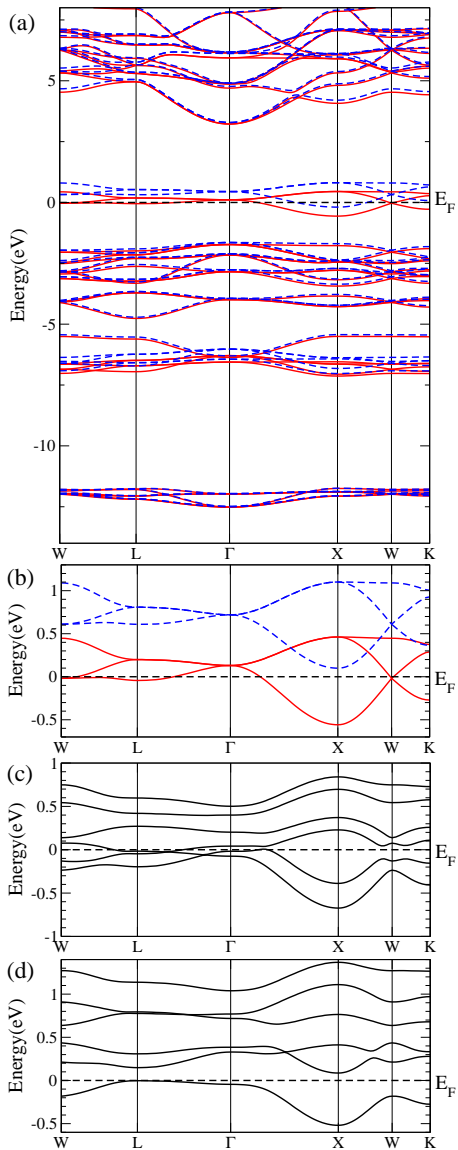}
  \caption{(Color online)
Band structures calculated for Ba$_2$NaOsO$_6$ using different methods. (a)
GGA, (b) GGA+U with $U_{eff}$ = 0.2 Ryd, (c) GGA+SOC, (d) GGA+SOC+U with
$U_{eff}$ = 0.2 Ryd. In (a) and (b) the solid and dashed lines refer to the
up- and down spin bands, respectively. In (d) the valence band top is
taken as the zero-energy point.} 
  \label{fig2}
\end{figure}

\begin{figure}[!hbp]
  \includegraphics[width=8.0cm]{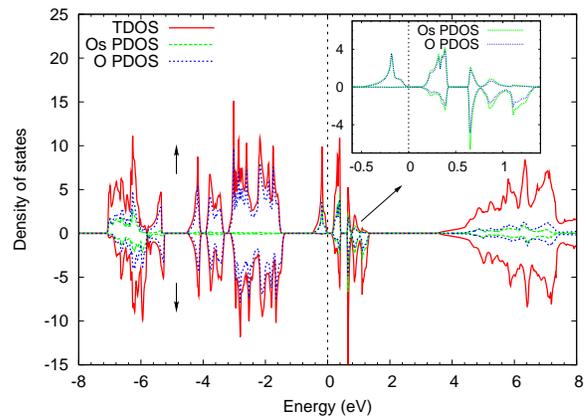}
  \caption{
    (Color online) DOS of Ba$_2$NaOsO$_6$ obtained from the GGA+SOC+U
    calculation. The inset shows the PDOS plots calculated for
    the Os and O atom contributions to the $t_{2g}$ bands.}
  \label{fig3}
\end{figure}

To gain insight into the puzzling electronic and magnetic properties
of Ba$_2$NaOsO$_6$, we carried out a first-principles density functional
theory (DFT) electronic structure study. In this Letter, we show that
the insulating property of Ba$_2$NaOsO$_6$ originates from a cooperative
effect of electron correlation and SOC. This cooperative effect is
responsible not only for the lack of structural distortion in
Ba$_2$NaOsO$_6$ but also for the low magnetic moment and the weak magnetic
anisotropy \cite{Erickson2006} of Ba$_2$NaOsO$_6$ in the FM state. 

Our first principles DFT electronic structure calculations were
performed by using the full-potential augmented plane waves plus local
orbital method as implemented in the WIEN2k code \cite{wien2k}. The
non-overlapping muffin-tin sphere radii of 2.50, 2.14, 1.86, and 1.65
au are used for the Ba, Na, Os and O atoms, respectively. The
expansion in spherical harmonics of the radial wave functions were
taken up to $l$ = 10. The value of $R_{MT}^{min}K_{max}$ was set to 7.0. The total
Brillouin zone was sampled with 125 k-points. For the
exchange-correlation energy functional, the generalized gradient
approximation (GGA) by Perdew, Burke and Ernzerhof \cite{Perdew1996} was employed. The
SOC was included on the basis of the second-variational method using
scalar relativistic wave functions \cite{Kunes2001}. The structural parameters of
Ba$_2$NaOsO$_6$ were taken from the experimental values \cite{Stitzer2002}. 

The spin-polarized GGA band structure calculated for the FM state of
Ba$_2$NaOsO$_6$, presented in Figure~\ref{fig2}(a), has the Fermi level crossing both
the up- and down-spin $t_{2g}$ bands. This does not agree with the fact
that Ba$_2$NaOsO$_6$ is an insulator \cite{Erickson2006}. At $\Gamma$ the three $t_{2g}$ bands are
degenerate. The overall width of the $t_{2g}$ bands is narrow
(approximately 1 eV) because the nearest-neighbor O$\cdots$O distance between
adjacent OsO$_6$ octahedra is long (3.216 \AA) compared with the van der
Waals distance of 3.04 \AA. The exchange splitting of Ba$_2$NaOsO$_6$ is about
0.34 eV, which is considerably smaller than typical values found for
3d magnetic oxides. Because the bandwidth is small compared with the
on-site repulsion U (approximately 3.3 eV), it was suggested
\cite{Erickson2006} that electron correlation is important in
Ba$_2$NaOsO$_6$. The failure of the traditional DFT in describing
strongly correlated systems is currently remedied by the DFT plus
on-site repulsion U  method \cite{LDA_U1,LDA_U2,LDA}. 
Thus, we employed the GGA+U method to see if the insulating
property of Ba$_2$NaOsO$_6$ can be explained. To avoid a double counting in
the non-spherical part of potential, we used $U_{eff} = U - J$ and omit the
multipolar terms proportional to $J$ in the added GGA+U potential. The
band structure calculated for the FM state of Ba$_2$NaOsO$_6$ using the
GGA+U method with $U_{eff}$ = 0.2 Ryd is shown in Figure~\ref{fig2}(b), which reveals
a larger exchange splitting (about 0.7 eV). However, there is no band
splitting at $\Gamma$, and the dispersion characteristics of the $t_{2g}$ bands
are almost the same as those of the GGA calculation. Our GGA+U
calculations with larger $U_{eff}$ values (up to 0.5 Ryd) did not change
the general picture described above. Since Os is a heavy element, SOC
is expected to play an important role in Ba$_2$NaOsO$_6$. To see the effect
of SOC on the electronic structure of Ba$_2$NaOsO$_6$, we performed GGA+SOC
calculations. Figure~\ref{fig2}(c) shows the band structure calculated for the FM
state using the GGA+SOC method with the spin quantization taken along
the [111] direction. The up- and down-spin $t_{2g}$ bands are both split
into three non-degenerate bands at $\Gamma$. However, these bands overlap
with each other leading to a metallic state for Ba$_2$NaOsO$_6$.

As described above, the GGA, GGA+U, and GGA+SOC methods all fail to
reproduce the insulating state for Ba$_2$NaOsO$_6$. Nevertheless, we note
that electron correlation enhances the exchange splitting, while SOC
splits the $t_{2g}$ bands. This suggests that a combined effect of
electron correlation and SOC might induce a band gap in both the up- and
down-spin $t_{2g}$ bands. Therefore, we carried out GGA+SOC+U calculations
for Ba$_2$NaOsO$_6$. The band structure calculated for the FM state with the
spin quantization along the [111] direction is presented in Figure~\ref{fig2}(d),
which shows that Ba$_2$NaOsO$_6$ has an insulating gap with the lowest-lying
down-spin $t_{2g}$ band lying above the Fermi level. The associated density
of states (DOS) calculated for Ba$_2$NaOsO$_6$ is shown in Figure~\ref{fig3}. The
partial DOS (PDOS) plots for the Os 5d and O 2p states reveal that the Os 5d
and the O 2p states contribute almost equally in the $t_{2g}$ bands. Though
not shown, the PDOS plots calculated for the Os 5$d_{xz}$, 5$d_{yz}$ and
5d$_{xy}$ orbitals show that these orbitals contribute equally to the $t_{2g}$
bands. The Os-O bonding bands of the $e_g$-symmetry occur well below the
Fermi level (i.e., in the energy region between $-7$ and $-5$ eV), which
indicates the presence of a strong covalent bonding in the Os-O
bonds.

So far, our calculations were performed for the FM state of
Ba$_2$NaOsO$_6$. To find if the FM state is the magnetic ground state, we
considered an A-type AFM state in which the spins have the FM ordering
within each sheet of Os$^{7+}$ ions parallel to the (001) plane, but have
the AFM ordering between adjacent sheets parallel to the (001)
plane. Our GGA+SOC+U calculation shows that the A-type AFM state is
considerably less stable than the FM state (by 267 meV/FU), which is
consistent with the observation that Ba$_2$NaOsO$_6$ undergoes an FM
ordering below 6.8 K \cite{Erickson2006}. In the remainder of this
work we will consider only the electronic structures calculated for
the FM state of Ba$_2$NaOsO$_6$. 

To examine the effect of the spin quantization direction on electronic
structure, we carried out GGA+SOC and GGA+SOC+U calculations for the
FM state of Ba$_2$NaOsO$_6$ with the spin quantization taken along the
[111], [110] and [100] directions. The stability dependence of the FM
state on the spin quantization direction is negligible in the GGA+SOC
calculations, but is not negligible in the GGA+SOC+U calculations. In
the latter calculations with $U_{eff} = 0.2$ Ryd, the FM states with the
[110] and [100] quantizations are less stable than that with the [111]
quantization by 1.9 and 16 meV per FU, respectively. The spin, orbital
and total moments calculated for the FM state of Ba$_2$NaOsO$_6$ using the
GGA+SOC+U method with $U_{eff} = 0.2$ Ryd are summarized in
Table I. For each of the [111], [110] and [100]
quantizations, the total moment per 
FU is calculated to be $\sim 0.65 \mu_B$, which is smaller than the spin-only
value of 1 $\mu_B$ due to the fact that the orbital moment of $\sim 0.33 \mu_B$ is
opposite to the spin moment in direction. Nevertheless, the calculated
values are still large compared with the zero-field moments ($\sim
0.2 \mu_B$/FU) 
determined from the magnetization study of Ba$_2$NaOsO$_6$ \cite{Stitzer2002,Erickson2006}.

The band gap opening at the Fermi level, the moment reduction and the
slight magnetocrystalline anisotropy in Ba$_2$NaOsO$_6$ can be accounted for
by considering the effect of SOC on the $t_{2g}$ orbitals of an Os$^{7+}$
ion. In the second variational approach for spin-orbit coupling \cite{Kunes2001},
the scalar-relativistic part of the Hamiltonian is diagonalized on a
basis adopted for each of the spin projections separately, and then
the full Hamiltonian matrix is constructed on the basis of the
eigenfunctions obtained in the first step. The spin-orbit part of the
Hamiltonian in the Os spheres is then given by 
\begin{equation}
\hat{H}_{so}=\lambda \hat {\mathbf {L}} \cdot \hat {\mathbf {S}},
\end{equation}
where the SOC constant $\lambda >0$ for the Os$^{7+}$ (d$^1$) ion with less
than half-filled $t_{2g}$ levels. 
With $\theta$  and $\phi$ as the azimuthal and polar angles of the
magnetization in the rectangular crystal coordinate system, the
$\hat{\mathbf {L}} \cdot \hat{\mathbf {S}}$ term is rewritten as 
\begin{equation}
  \begin{array}{lll}
  \hat{\mathbf {L}} \cdot \hat{\mathbf {S}} &=&
  \hat{S}_z (\hat{L}_z \cos \theta + \frac{1}{2} \hat{L}_+ e^{-i \phi} \sin \theta  +\frac{1}{2}
  \hat{L}_- e^{i \phi} \sin \theta ) \\
&&+\frac{1}{2}\hat{S}_+ (-\hat{L}_z \sin \theta -  \hat{L}_+ e^{-i \phi} \sin^2 \frac{\theta}{2}  +
  \hat{L}_- e^{i \phi} \cos^2 \frac{\theta}{2} ) \\
  &&+\frac{1}{2}\hat{S}_- (-\hat{L}_z \sin \theta -  \hat{L}_+ e^{-i \phi} \cos^2
  \frac{\theta}{2}  +
  \hat{L}_- e^{i \phi} \sin^2 \frac{\theta}{2} ).
  \end{array}  
  \label{eq2}
\end{equation}
Since the up- and down-spin $t_{2g}$ bands are separated due to the
exchange splitting, one can neglect interactions between the up- and
down-spin states under the SOC to a first order approximation. This
allows one to consider only the up-spin $t_{2g}$ bands using the degenerate
perturbation theory, which requires the construction of the matrix
elements $\langle i | \hat{H}_{so} |j \rangle$ 
($i, j = d_{xy}, d_{yz}, d_{xz}$) \cite{Dai2005}. In this case, only the
operators of the first line of Eq.~\ref{eq2} give rise to nonzero matrix
elements. In evaluating these matrix elements, it is convenient to
rewrite the angular parts of the $d_{xy}$, $d_{yz}$ and $d_{xz}$ orbitals in terms of
the spherical harmonics as 
\begin{equation}
  \begin{array}{l}
    d_{xy}=\frac{-i}{\sqrt{2}}(Y_{2}^{2} - Y_{2}^{-2}) \\
    d_{yz}=\frac{i}{\sqrt{2}} (Y_{2}^{1} + Y_{2}^{-1}) \\
    d_{xz}= \frac{-1}{\sqrt{2}} (Y_{2}^{1} - Y_{2}^{-1}) .
  \end{array}
\end{equation}
Using these functions, the matrix representation  $\langle i | \hat{H}_{so}
|j \rangle$ is found as 
\begin{equation}
  i\hbar \lambda/2 \left(
  \begin{array}{ccc}
     0 &  \sin \theta \sin \phi & - \sin \theta
    \cos \phi \\
     - \sin \theta \sin \phi & 0 &  \cos \theta  \\
      \sin \theta  \cos \phi & - \cos \theta & 0
  \end{array}\right).
\end{equation}
Upon diagonalizing this matrix, we obtain the eigenvalues of the three
spin-orbit coupled states, namely,
$E_1=-\hbar \lambda /2 $, $E_2=0 $, $E_3= \hbar \lambda /2 $. The
associated eigenfunctions $\Psi_1, \Psi_2, \Psi_3$ are given by  
\begin{equation}
  \begin{array}{lll}
    \Psi_1 &=& \frac{\sqrt {2}}{2} [\sin \theta d_{xy} +(i \sin \phi - \cos \theta  \cos
    \phi) d_{yz} \\
    & &- ( i \cos \phi     + \cos \theta \sin \phi) d_{xz} ] \\
    \Psi_2 &=& \frac{\sqrt {2}}{2} [ \sin \theta d_{xy} - (i \sin \phi + \cos \theta \cos
    \phi) d_{yz} \\ 
    & &+ (i \cos \phi - \cos \theta \sin \phi) d_{xz} ] \\
    \Psi_3 &=& \cos \theta d_{xy} + \sin \theta \cos \phi  d_{yz} +
    \sin \theta  \sin \phi d_{xz} .
  \end{array}
  \label{eq5}
\end{equation}
For these three states $\Psi_1$, $\Psi_2$, and $\Psi_3$, the orbital moments $L$ along the
spin quantization direction are $-1 \mu_B$, 0 and $1 \mu_B$, respectively,
according to their eigenvalues and the SOC operator  
$\hat{H}_{so}=\lambda \hat{\mathbf {L}} \cdot \hat{\mathbf {S}}$ 
given that S = 1/2. Since the spin-orbit coupled state $\Psi_1$ is
occupied, we obtain a negative orbital moment. The moment of $-0.35
\mu_B$
from the GGA+SOC+U calculation is considerably smaller in magnitude
than $-1 \mu_B$. This is not surprising because the above analysis
neglected the fact that the $t_{2g}$ orbitals of an OsO$_6$ octahedron are not
pure 5$d_{xz}$, 5$d_{yz}$ and 5$d_{xy}$ orbitals of the Os atom, but are given by
their linear combinations with the 2p orbitals of the surrounding O
atoms. Given that the Os 5d and the O 2p states contribute almost
equally in the $t_{2g}$ bands (Figure~\ref{fig3}), an orbital moment of
approximately $-0.5  \mu_B$ should be expected from the occupation of
the up-spin $t_{2g}$ band associated with $\Psi_1$.

The above analysis indicates that SOC splits the $t_{2g}$ bands into three
subbands regardless of the direction of the magnetization. However,
SOC does not affect the exchange splitting. It is the on-site
repulsion that enhances the exchange splitting and increases the
energy separation between filled and empty bands within each spin
channel. That is, a cooperative effect of electron correlation and SOC
is essential in opening a band gap at the Fermi level for
Ba$_2$NaOsO$_6$. Eq.~\ref{eq5} shows that the contributions of the
$d_{xy}$, $d_{yz}$ and $d_{xz}$ orbitals to the spin-orbit coupled
state $\Psi_1$ depend on the direction of the magnetization (i.e.,
$\theta = 90 ^{\circ}$ and $ \phi = 0 ^{\circ}$ for [100]; 
$\theta = 90 ^{\circ}$ and 
$ \phi = 45 ^{\circ}$ for [110];  $\theta = \arccos(\sqrt{3}/3)$ and $ \phi
= 45 ^{\circ}$ for [111]).  This is responsible for the weak magnetic anisotropy observed for
Ba$_2$NaOsO$_6$ \cite{Erickson2006}. The three d-orbitals contribute
equally to the state $\Psi_1$ 
for the [111] spin quantization, but unequally for the [110] and [100]
spin quantizations. The inter-octahedron hopping integral is nonzero
along the directions of nearest-neighbor OsO$_6$ octahedra (e.g., [111]
and [110]) but is practically zero along the directions of
next-nearest-neighbor OsO$_6$ octahedra (e.g., [100]). For the [100] spin
quantization, the $\Psi_1$ level does not provide hopping along the [011]
direction because it has no $d_{yz}$ orbital contribution. Consequently,
the spin-orbit coupled band associated with $\Psi_1$ should be higher in
energy for the [100] quantization than for the [111] and [110]
quantizations. This in part explains why the [100] quantization leads
to a higher electronic energy than do the [111] and [110]
quantizations.

In summary, the insulating behavior of Ba$_2$NaOsO$_6$ is caused by a novel
cooperative effect of electron correlation and SOC, which opens a band
gap at the Fermi level and hence removes a driving force for orbital
ordering of the orbitally degenerate Os$^{7+}$(d$^1$) ions. The small magnetic
moment of Ba$_2$NaOsO$_6$ arises from the fact that the occupied
spin-orbit-coupled up-spin $t_{2g}$ band gives rise to an orbital moment
that is in opposite direction to the spin moment.

% Acknowledgement.
The research was supported by the
Office of Basic Energy Sciences, Division of Materials
Sciences, U. S. Department of Energy, under Grant
No. DE-FG02-86ER45259.

%\clearpage

%\clearpage


\begin{thebibliography}{99}
\bibitem{LDA}Our calculations using the local density approximation led
essentially to the same results except that a larger $U_{eff}$ ($ > 0.21$
Ryd) is needed to open an insulating gap.
\end{thebibliography}
\end{document}